\documentclass[aps,prl,amsmath,amssymb,reprint,numeric]{revtex4-1}
\usepackage{graphicx}
\usepackage{subfigure}
\usepackage{color}
\usepackage{amsmath}
\usepackage{bm}
\usepackage[normalem]{ulem}
\usepackage{multirow}
\usepackage{float}

\begin{document}

\title{A Microscopic Theory of Softness in Supercooled Liqiuds}
\author{Manoj Kumar Nandi}
\address{\textit{Polymer Science and Engineering Division, CSIR-National Chemical Laboratory, Pune-411008, India}}
\author{Sarika Maitra Bhattacharyya}
\email{mb.sarika@ncl.res.in}
\address{\textit{Polymer Science and Engineering Division, CSIR-National Chemical Laboratory, Pune-411008, India}}

\begin{abstract}

We introduce a new measure of the structure of a liquid which is the softness of the mean-field potential developed by us earlier. We find that this softness is sensitive to small changes in the structure. We then study its correlation with the supercooled liquid dynamics. The study involves a wide range of liquids (fragile, strong, attractive, repulsive, and active) and predicts some universal behaviours like the softness is linearly proportional to the temperature and inversely proportional to the activation barrier of the dynamics with system dependent proportionality constants. We write down a master equation between the dynamics and the softness parameter and show that indeed the dynamics when scaled by the temperature and system dependent parameters show a data collapse when plotted against softness. The dynamics of fragile liquids show a strong softness dependence whereas that of strong liquids show a  much weaker softness dependence. We also connect the present study with the earlier studies of softness involving machine learning (ML) thus providing a theoretical framework for understanding the ML results.

\end{abstract}

\maketitle

\section{Introduction}

In liquid state theory, the structure of a system is the foundation on which the theories are built \cite{ry_form,hansen_mcdonald,gotze}. However in a series of paper, Berthier and Tarjus investigated the role of structure in the dynamics showing that at low temperatures although the pair structures of the liquids where particles are interacting via the Lennard-Jones(LJ) and its repulsive counterpart the WCA \cite{wca_model} potentials are quite similar, the dynamics are quite different\cite{kobandersenLJ,berthier_terjus_PRL_103_170601_2009, PRE_82_031502_2010,EPJE_34_1_2011, JCP_134_214503_2011}. Thus questioning the role of structure in the dynamics and the validity of the liquid state theories like the mode coupling theory (MCT) \cite{gotze,gotze_condmat} and the density functional theory \cite{schweizer2003,schweizer2005} specially in the supercooled liquid regime.

Subsequent work involving some of us has shown that the pair configurational entropy, described by just the pair structure, $S_{C2}$ of the LJ and WCA systems are quite different\cite{role_pair_configuration}. Thus the dynamics described by the pair configurational entropy, via the Adam-Gibbs \cite{adam_gibbs} relation are also apart. This clearly showed that a small difference in structure can lead to a large difference in dynamics. Later Schweizer and coworkers proposed a new microscopic theory where they worked with the real forces instead of the projected one used in MCT \cite{schweizer_prl2015} and used the structure to differentiate between the dynamics of the LJ and WCA systems.     

Further study around this newly described function, $S_{C2}$ revealed some interesting but puzzling results \cite{unraveling}. Like the activated regime described by Adam-Gibbs relation \cite{adam_gibbs} overlaps with the MCT regime and the configurational entropy follows a power law behaviour appearing to vanish at the MCT transition temperature,$T_{C}$ \cite{unraveling}. The genesis of these was found to be the vanishing of the pair configurational entropy at $T_{C}$.   The only possible connection between $S_{C2}$ related to an activation process and non-activated MCT dynamics is the structure, implying that the structure has the information of the dynamical transition.   

To validate this hypothesis, Nandi {\it et al} developed a microscopic mean-field theory \cite{role_pair_correlation}. In this theory, the mean field potential which is essentially a caging potential of a particle was described in terms of the pair structure of the liquid and it was shown that for a large number of systems the escape dynamics of the particle from this trapping potential follows the MCT power law behaviour and  diverges at the dynamical transition temperature. Thus confirming that the information of the dynamical transition temperature is embedded in the pair structure of the liquid \cite{role_pair_correlation}. 

In a recent study, the role of structure in the dynamics was revisited \cite{atractive_truncated_Landes}. This study was built on earlier studies  where a `softness' parameter was defined, which is the weighted integral of the pair correlation function \cite{cubuk,Schoenholz_nature,Schoenholz263}. The weights were obtained from the correlation of the structure and the dynamics using machine learning (ML) techniques. According to these studies, below the onset temperature of glassy dynamics \cite{sastry_debenedetti}, the dynamics is controlled by the softness of the system  \cite{entropy_onset}. The more recent study showed that the `softness' parameter of the LJ and WCA systems are different which eventually leads to the difference in their dynamics \cite{atractive_truncated_Landes}. Despite the success of the ML study in connecting the softness parameter to the dynamics, the connection between the softness parameter and the structure is not clear. 

The main objective of this study is to understand the correlation between the structure and dynamics, and the question boils down to what measure of the structure to use. Towards this goal, we define a new {\it microscopic measure of the structure}, which is the softness (slope) of the mean-field caging potential developed by us earlier \cite{role_pair_correlation}. We find that the softness and its temperature dependence varies with the system. The study reveals that the activation barrier for the dynamics is inversely proportional to the softness with system dependent proportionality constants. Using this information we can write a master equation between the dynamics and softness and show that for a wide range of systems the simulated dynamics (scaled by these parameters) follow this master equation and show a data collapse when plotted against the softness.

In the spirit of the mean-field theory developed earlier, we can map the dynamics of a set of  interacting particles in terms of dynamics of a set of non-interacting particles in a potential $\Phi(r)$\cite{role_pair_correlation,indranil} (See SI for details \cite{supplemental}). Like earlier studies of density functional theory \cite{wolynesspra,xiawolyness,archer,hopkins2010} we study the properties of the scaled potential $\beta\Phi(r)$
\begin{eqnarray}
\beta\Phi(r)=-\int \frac{d{\bf{q}}}{(2\pi)^3}\sum_{\alpha\beta} C_{\alpha\beta}(q)\sqrt{x_\alpha x_\beta}(S_{\alpha\beta}(q)-\delta_{\alpha\beta})e^{-q^2r^2/6}.
\label{mfpt_pot_binary}
\end{eqnarray}
Where $x_{\alpha/\beta}$ represents the fraction of particles of type $\alpha/\beta$ in the binary mixture. In the above expression $S_{\alpha\beta}({\bf q})=\frac{1}{\sqrt{N_{\alpha}N_\beta}}\sum_{i=1}^{N_\alpha}\sum_{j=1}^{N_\beta} exp[-i{\bf q}.({\bf r}_i^{\alpha}-{\bf r}_j^{\beta})]$ and ${\bf{C(q)=1-S^{-1}(q)}}$, where $S_{\alpha\beta}(q)$ is the partial structure factor of the liquid and ${\bf{C}}$, ${\bf{1}}$ and ${\bf{S(q)}}$ are in matrix form.
This potential is an effective caging potential of each particle.

\begin{figure}[ht]
	\centering
	\subfigure{
\includegraphics[width=0.44\textwidth]{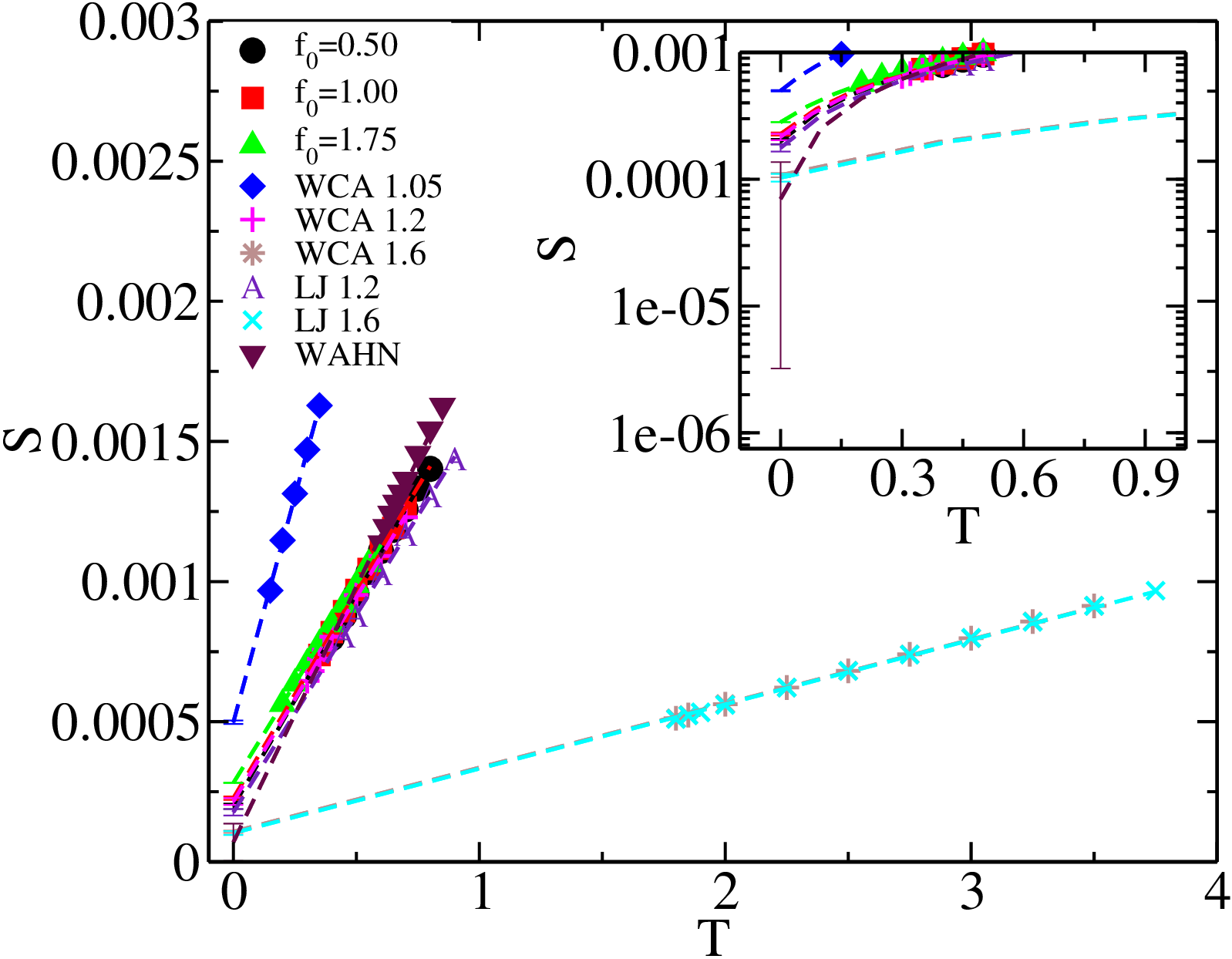}}

	\caption{\it{Softness, $S$ against temperature for all the systems studied here \cite{system_details}. Softness is calculated by fitting the mfpt potential,( Eq.\ref{mfpt_pot_binary}), to a harmonic form near the minimum. The density of the WCA and the LJ systems are mentioned next to them and WAHN stands for Wahnstr\"om model \cite{wahn_model}. $f_0=0.5,1.00,1.75$, are LJ 1.2 systems with activity (self-propulsion force) $f_{0}$ (See reference \cite{system_details} for details). (Inset) We have zoomed in near the origin and plotted the y-axis in logarithmic scale to show the error bars.}}
	\label{fig.3a}
\end{figure} 

To quantify the softness of the potential we fit the potential near r=0 to a harmonic form, $\beta \Phi(r)=\beta \Phi(r=0)+\frac{1}{2}r^2/S$, where $\Phi(r=0)$ is the value of the potential at the minima and $S$, presented without indices, represents the softness of the potential. As described by Eq.\ref{mfpt_pot_binary} as long as we have the information of the partial structure factors, $S_{\alpha\beta}$, we can calculate the caging potential and its softness. Thus the softness can be calculated a priori in theory, simulations and experimental studies. In this study, we use the simulated values of the partial structure factors. For all the systems studied here, we plot the calculated softness against temperature (Fig.\ref{fig.3a}) \cite{system_details}(System details are also given in the SI). The systems studied here not only include a wide range of softness but also cover a wide variety in terms of their temperature dependence.  
 Despite this variation, we observe some universal phenomena. For all the systems with a decrease in temperature, the softness decreases linearly and does not appear to vanish even at zero temperature (except for the WAHN model where the error bar does include nonzero but very small values). With a decrease in temperature, the structure becomes more well defined leading to an increase in the depth of the caging potential (See SI \cite{supplemental}) and a decrease in softness. The LJ and the WCA systems are studied at more than one density. For these systems, we find that with an increase in density the softness decreases which can be justified as at higher densities the systems are expected to be better packed. Also, the rate of decrease of softness with temperature is lower at higher densities suggesting that at higher densities the structure shows a weaker variation with temperature. For the active systems, the softness increases with the activity. This is commensurate with the earlier observation that with an increase in activity the cage size increases \cite{active_system}. 

\begin{figure*}[ht]
	\centering
	\subfigure{
	\includegraphics[width=0.42\textwidth]{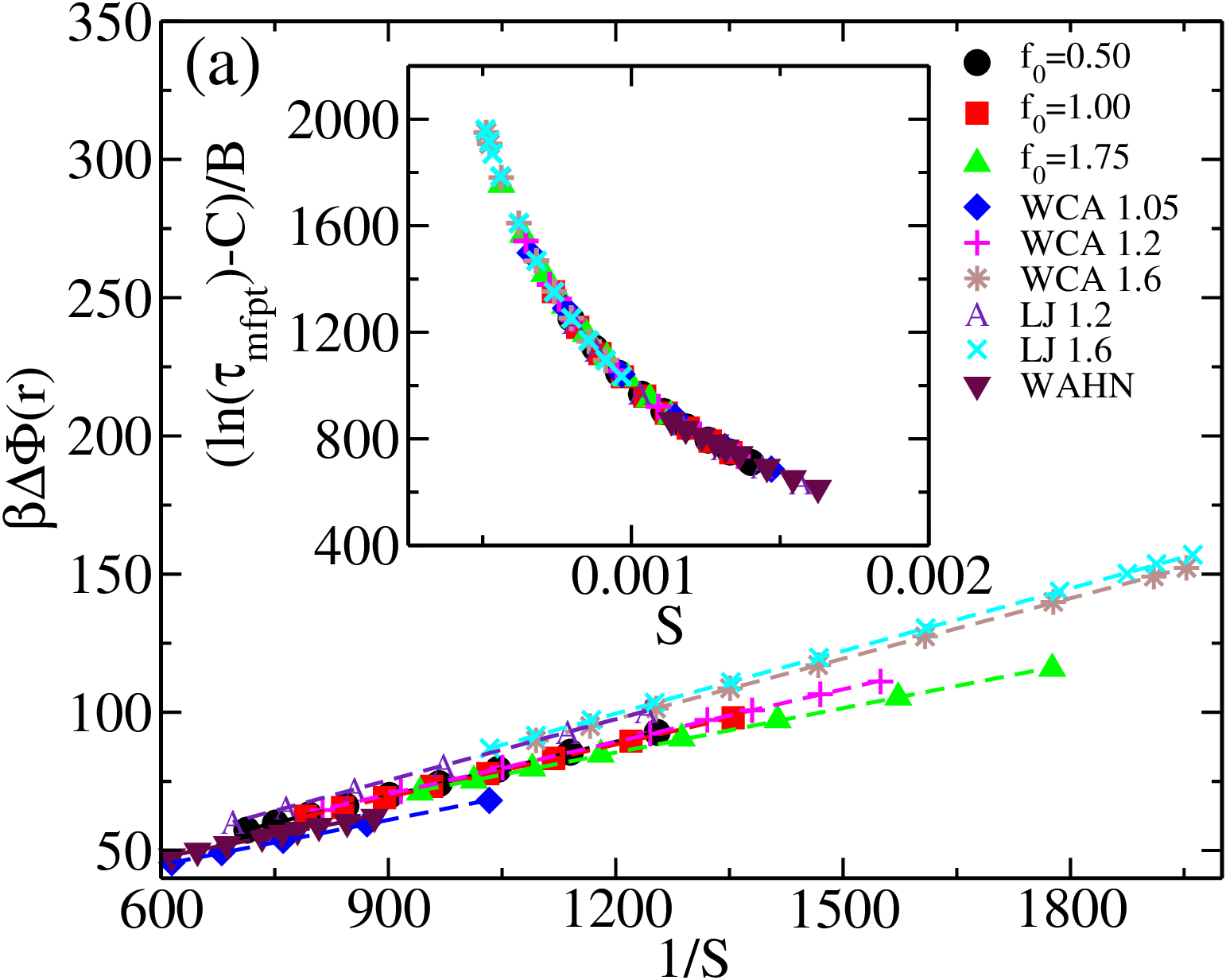}}
	\subfigure{
    	\includegraphics[width=0.44\textwidth]{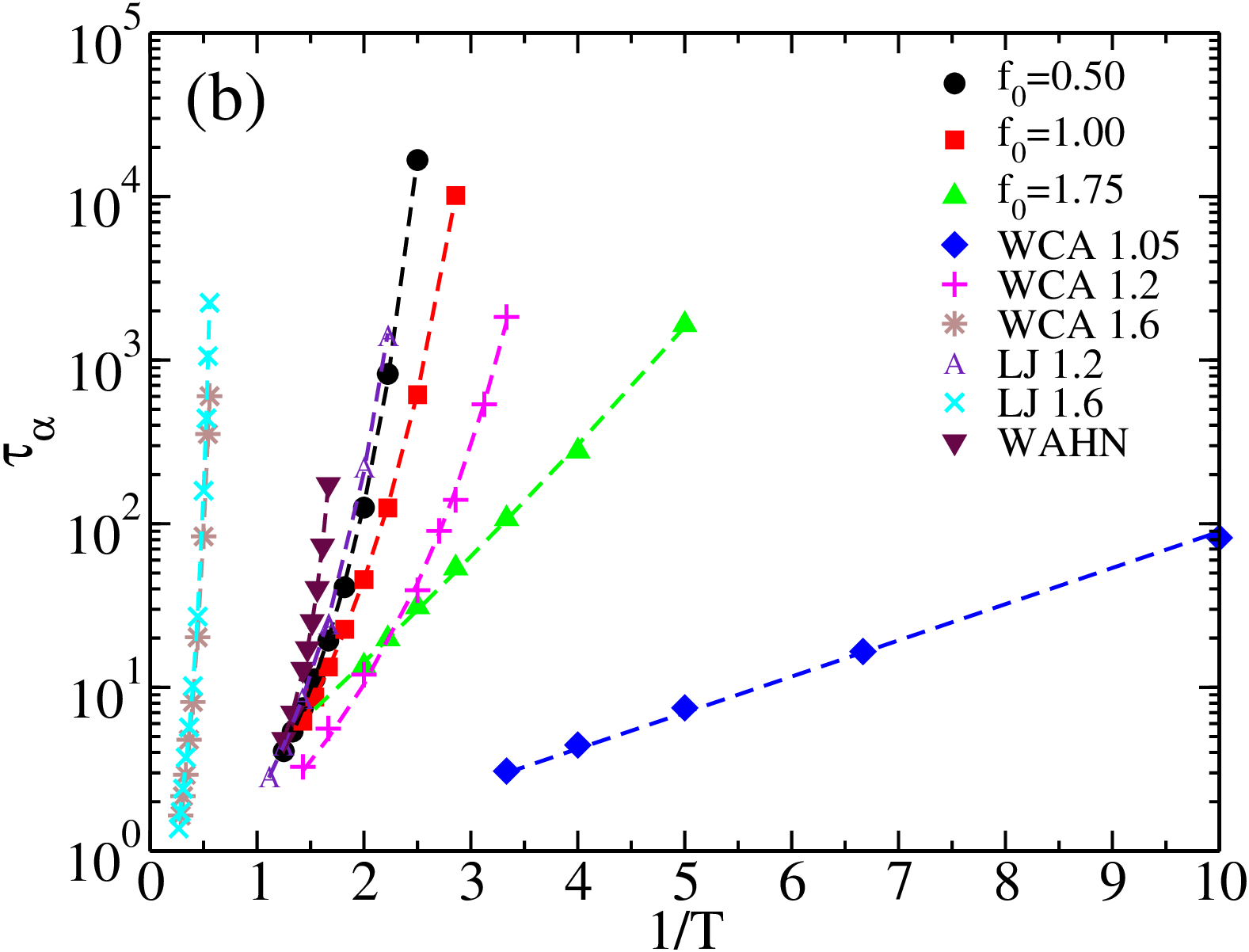}}
	\subfigure{
		\includegraphics[width=0.44\textwidth]{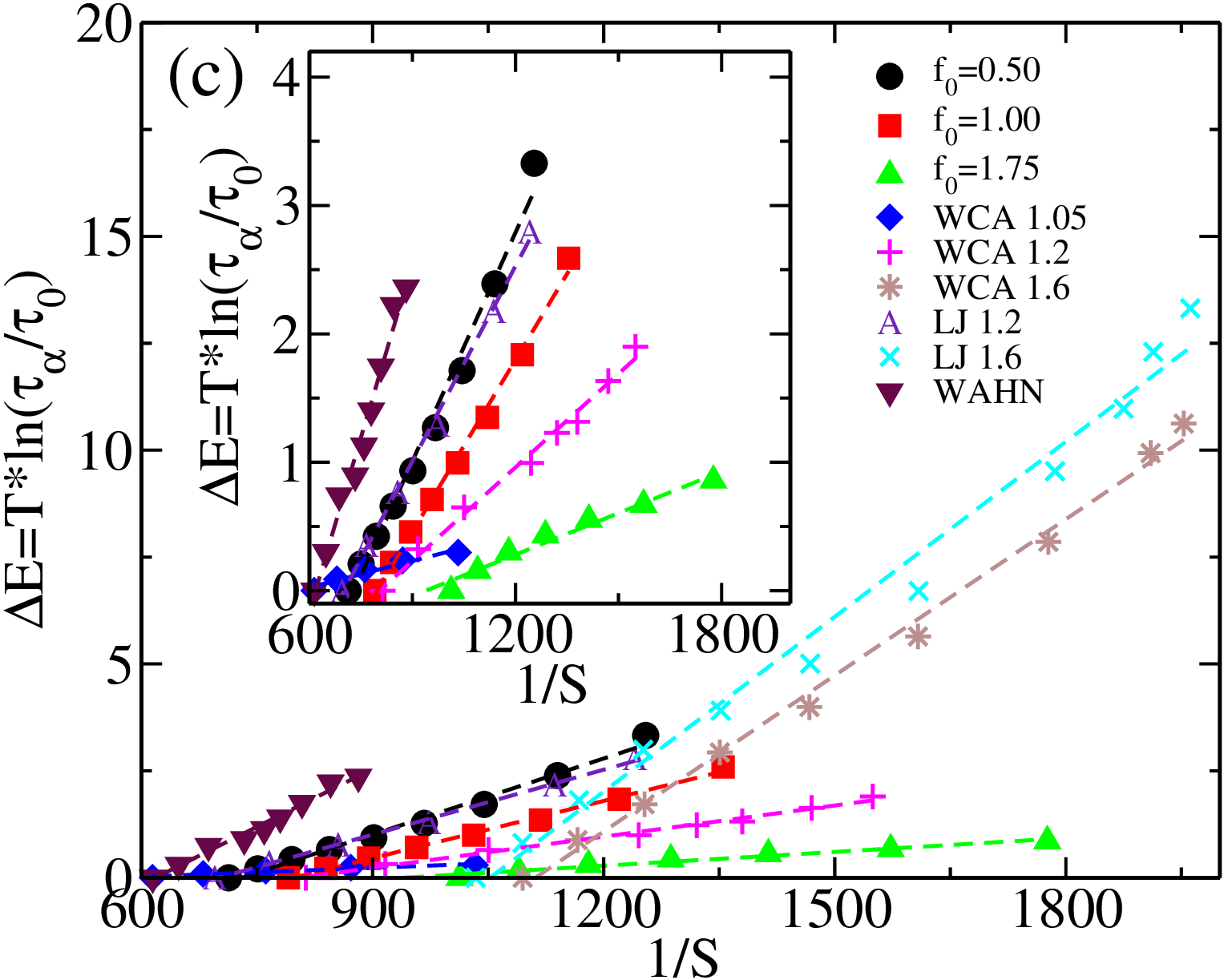}}
		\subfigure{
			\includegraphics[width=0.44\textwidth]{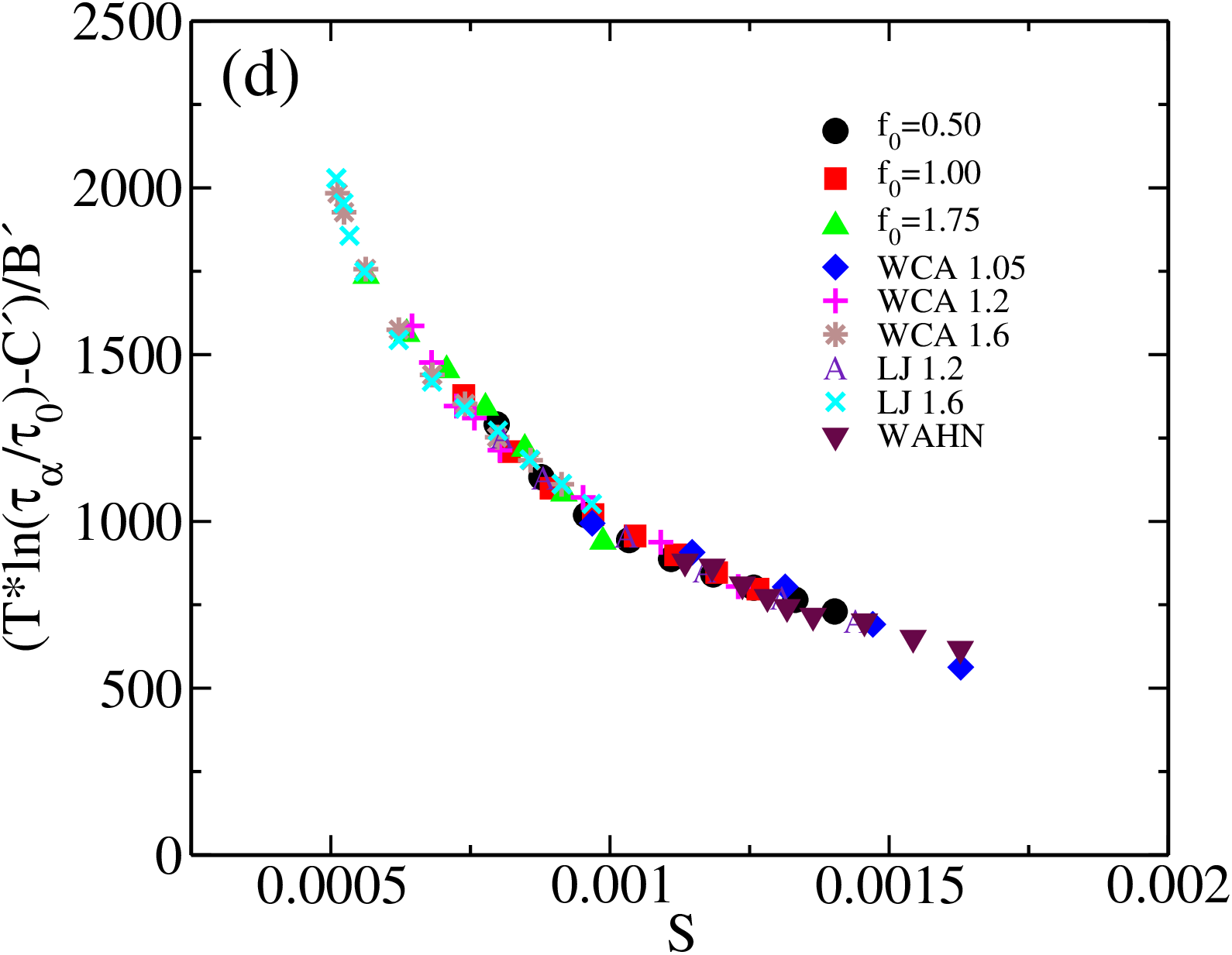}}
			
	\caption{\it{(a) The potential barrier $\beta\Delta \Phi(r)=-\beta\Phi(r=0)$ (Eq.\ref{mfpt_pot_binary})  against inverse of softness. The plot shows linear behaviour. (inset) $\frac{\ln(\tau_{mfpt})-C}{B}$ shows a master curve for all the systems when plotted against softness, $S$. Here C and B are the system dependent parameters obtained from the linear fit of $\ln(\tau_{mfpt})$ against 1/S.
	(b) Simulated relaxation time $\tau_{\alpha}$ as a function of inverse temperature \cite{system_details}. The systems cover a wide range of fragility\cite{fragility_value}.
	(c) $\Delta E=T*\ln(\tau_{\alpha}/\tau_{0})$ plotted against the inverse of softness. For each system the data can be described by a linear fit with system dependent parameters. The inset zooms on to some of the systems.
	(d) $\frac{T*\ln(\tau_\alpha/\tau_0)-C'}{B'}$ vs. softness, $S$ shows a master curve for all the systems which is a proof of validity of Eq.\ref{strong_fragile}. Here C' and B' are obtained from Fig\ref{delphi_plots}(c). }}

	\label{delphi_plots}
\end{figure*}

The question that we ask next is how the softness affects the dynamics of a system. However, before we study the role of the softness in the actual dynamics of the system as obtained from the simulation studies (See SI for brief description \cite{supplemental}, which includes Ref. [\onlinecite{overlap_function}]) we try to get some insight from the correlation of softness and the mean first passage time (mfpt) dynamics. The mfpt dynamics, $\tau_{mfpt}$ is the time required for a particle to escape from the trapping potential $\beta\Phi(r)$. This mfpt dynamics can be expressed as,\cite{zwanzigPNAS,role_pair_correlation} . 
\begin{eqnarray}
 \tau_{mfpt}=\frac{1}{D_0}\int_0^{r_{max}} e^{\beta\Phi(y)}dy\int_0^ye^{-\beta\Phi(z)}dz.
 \label{taumfpt}
\end{eqnarray}
Where $D_0=k_BT/\zeta$ and $r_{max}$ is the range of the caging potential $\Phi(r)$ and $\zeta$ is the value of friction. Without loss of generality, we consider $D_{0}=1$, which is related to the high temperature barrier less process \cite{zwanzig}. It has been discussed earlier that in certain limit Eq.\ref{taumfpt} reduces to the expression of Kramers barrier crossing dynamics \cite{zwanzig,indranil}(See SI for details \cite{supplemental}). 
$\tau_{mfpt} \simeq \tau_{mfpt0}e^{\beta\Delta \Phi(T)}$ where the barrier height $\Delta \Phi= (\Phi_{max}-\Phi_{min})=-\Phi(r=0)$. Since the potential is monotonic, which has a minima at $r=0$ and disappears at '$r=r_{max}$', this barrier height is the depth of the potential (see SI for the plot of the potential, \cite{supplemental}). If we now assume a harmonic potential, we can write $1/S=2\beta\Delta\Phi/{r^2_{max}}$. Since $r_{max}$ (range of the potential) does not vary much with temperature (See SI for details \cite{supplemental}) the softness is expected to decrease inversely with the depth. As shown in Fig.\ref{delphi_plots}(a), for all the systems $ \beta \Delta \Phi$ is indeed linearly proportional to $1/S$.  We can write $\beta\Delta \Phi=ln(\tau_{mfpt}/\tau_{mfpt0})=C+B/S$ where $C$ and $B$ are system dependent parameters. In the inset of Fig.\ref{delphi_plots}(a) we plot $\frac{\ln(\tau_{mfpt})-C}{B}$ against softness which shows a master plot ($\tau_{mfpt0}=1/D_{0}=1$). Note that Kramer's relation between $\tau_{mfpt}$ and $\Delta \Phi$ is approximate, thus we obtain $C$ and $B$ from {$ln(\tau_{mfpt}/\tau_{0})$} vs $1/S$ fit. The master plot in the inset of Fig \ref{delphi_plots} (a) between $(ln(\tau_{mfpt})-C)/B$ vs. S tells us that when we scale the dynamics with these system dependent parameters we find a unique relation between the dynamics and the softness.

Motivated by these results we now address the main objective of this study i.e the correlation between the simulated dynamics ($\tau_{\alpha}$)  and the mean-field softness \cite{system_details,role_pair_correlation}. The temperature dependence of the $\tau_{\alpha}$ can be written in the form of activated dynamics $\tau_{\alpha}={\tau_0}exp[\frac{\Delta E}{T}]$ where $\tau_0$ is the value of $\tau_\alpha$ at the onset temperature of glassy dynamics \cite{entropy_onset} and $\Delta E$ is the temperature dependent barrier height. In Fig.\ref{delphi_plots}(b) we plot $\tau_{\alpha}$ against 1/T showing a large variation in terms of the temperature dependence of the dynamics. Using the insight obtained from the mfpt study, in Fig.\ref{delphi_plots}(c) we plot $\Delta E=T*ln(\tau_{\alpha}/\tau_{0})$ against $1/S$ which shows a linear behaviour for all the systems. $\Delta E=C'+B'/S$ where $C'$ and $B'$ are system dependent parameters obtained from the linear fit in Fig.\ref{delphi_plots}(c). We can now write 
\begin{equation}
\tau_{\alpha}={\tau_0}exp\Big[\frac{C'}{T}+\frac{B'}{TS}\Big].
\label{strong_fragile}
\end{equation}
In Fig.\ref{delphi_plots}(d) we plot $\frac{T*\ln(\tau_{\alpha}/\tau_0)-C'}{B'}$ against $S$  which shows a master plot and a proof of validity of our proposed master equation (Eq.\ref{strong_fragile}). The most important result is that the collapse of the seemingly disparate data does not require any adjustable parameters. The analysis thus suggests that there is a unique relationship between the dynamics and the softness (Eq.\ref{strong_fragile}) which is invariant of the type of interaction potential (attractive and repulsive) and also the kind of glass former (fragile and strong). However, a basic assumption in this theoretical model is that the caging potential is stable. Studies have shown that the local structure which gives rise to the caging potential is stable only below the onset of the glassy dynamics \cite{morse,berthier_ozawa_coslovich}. Thus this correlation between the dynamics and softness will be valid only below this temperature range an observation also made in the ML studies \cite{Schoenholz263}.

 \begin{figure}[ht]
	\centering
		\includegraphics[width=0.4\textwidth]{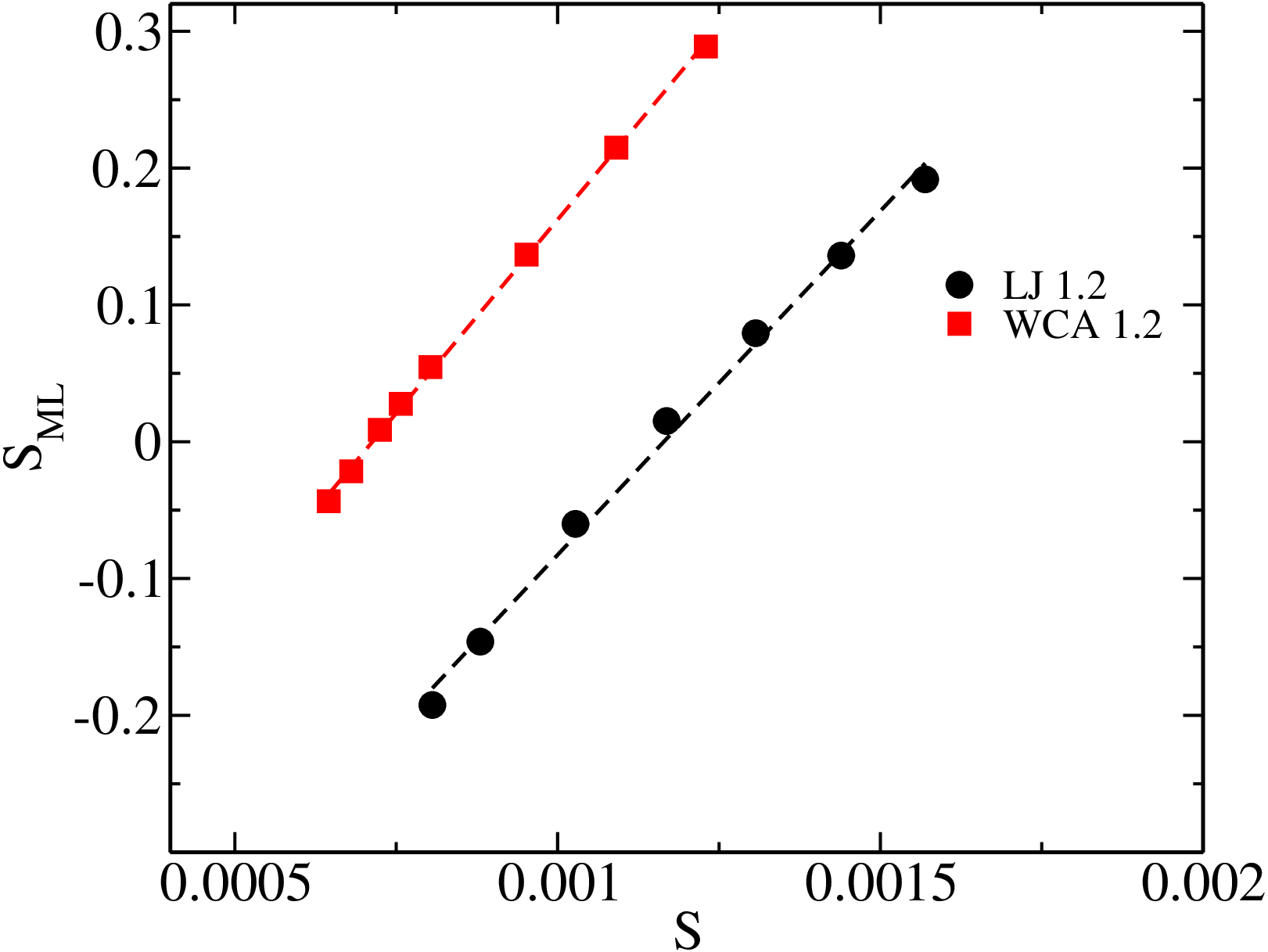}

		\caption{\it{The softness parameter obtained in ML study,$S_{ML}$ (Ref.\cite{atractive_truncated_Landes}) vs. the present softness parameter S. A linear correlation exists between them with system independent slope. }}
	\label{softness}
\end{figure}

The temperature dependence of the relaxation time (Eq.\ref{strong_fragile}) and the data collapse can be compared with the existing well known Vogel-Fulcher-Tamman (VFT) \cite{vft_relation_dyre}, Parabolic and other forms \cite{chandler,ELmatad_chandler}.  The present theory although predicts super Arrhenius behavior but unlike the VFT and other exponential forms \cite{berthier_colloid} and akin to the Parabolic form it does not predict a divergence of the relaxation time at a finite temperature. Also note that unlike the VFT form and again similar to the Parabolic \cite{ELmatad_chandler} and other forms \cite{berthier_colloid}, the fitting parameters are less sensitive to the temperature regime suggesting that they are indeed related to the material properties. The most important fact is that like the Shoving \cite{dyre} and KSZ models \cite{allessio, ksz_prr} the present form connects the local structural properties to the dynamics.

Our study reveals that although the structure of the LJ and WCA systems are similar, for the LJ system the energy barrier, $\Delta E$ has a stronger dependence on softness (the slope $B'$ in Fig.\ref{delphi_plots}(c) is larger). This result is consistent with the earlier predictions that the role of attractive force is to increases the energy barrier \cite{atractive_truncated_Landes,ELmatad_chandler}.
$B'$ appears to have a correlation with fragility \cite{fragility_value} (Fragility values are also tabulated in SI \cite{supplemental}).
For strong liquids, $B' \simeq 0$, on the other hand for fragile systems $B'$ is large. Thus for strong liquids in Eq.\ref{strong_fragile}, the softness independent first term dominates suggesting that the softness does not influence the dynamics. For fragile liquids, the softness dependent second term dominates and since the softness is dependent on the temperature it gives rise to the super Arrhenius behaviour.

Similar to the KSZ model \cite{allessio, ksz_prr} we find that for strong liquids the softness is generally high (WCA at $\rho=1.05$). However, in the KSZ model, the softness is almost temperature independent whereas in our analysis it varies with temperature. Nonetheless as mentioned above, for strong liquids the variation of the softness does not affect the dynamics. Thus although the softness of the active system, ($f_{0}=1.75$) and the Wahnstr\"om model (WAHN) \cite{system_details} are in a similar range, the former is a strong liquid whereas the latter is a fragile liquid.
The temperature dependence of the softness along with $B'$ is related to the temperature dependence of the energy barrier and thus to the fragility. For example, the WAHN model and the WCA 1.6 and LJ 1.6 systems have similar values of $B'$ but the temperature dependence of the softness is stronger for the more fragile WAHN model. 

According to the Isomorph theory for a strongly correlated system, along an isomorphic curve, the dynamics and the structure remains the same \cite{dyre_isomorph1,dyre_isomorph2,dyre_isomorph3}. Since the softness in our theory is directly related to the structure we expect that along the isomorphic curves the softness and dynamics will remain the same. This conjecture will be checked in future studies.
In the present theory we although define a caging potential, we cannot define a cage size and also our simulation results of the structure, required to describe the caging potential are above the dynamical transition temperature. Thus it is not possible to compare our results with a recent higher dimensional study of the cage size and its density/temperature dependence \cite{kundu}.

Finally, we compare the present work with the ML studies \cite{Schoenholz_nature,atractive_truncated_Landes}. In the ML study softness is measured as a distance from a hyperplane of multiple structure functions describing the local structure. The hyperplane is created using training sets having information of both the mobility and these structure functions and is expected to differentiate between the fast and the slow particles. The distance from the hyperplane and thus the softness can be both positive and negative. Thus in ML studies, the softness is a relative quantity whereas the softness in the present analysis is an absolute value in the range of only positive numbers.
In Fig.\ref{softness} we plot the average value of the ML softness, $S_{ML}$ for the LJ and WCA systems (taken from Ref \cite{atractive_truncated_Landes}) against the corresponding softness parameter $S$ calculated in the present study.  Despite the difference in the two methodologies we surprisingly find a linear behaviour suggesting that these two apparently different measures of softness are correlated over the whole temperature regime. Note that in ML studies the connection between the softness and the structure is not clear whereas in our study we calculate softness from the structure. Thus right now we do not have a complete understanding of this correlation except for the fact that both $S_{ML}$ and $S$ are related to the structure of the liquid and probably their correlation can be used to connect $S_{ML}$ to the structure via $S$. 

Here we introduce a new measure of the structure of a liquid which is the softness of the mean field potential developed by us earlier \cite{role_pair_correlation}. The study involves a wide range of systems showing different degrees of fragility. For all the systems the softness shows a linear dependence with the temperature where the slope is system dependent. The energy barrier related to the dynamics varies linearly with inverse softness. For strong liquids, the energy barrier shows a weak dependence on softness whereas for fragile liquids it shows a strong variation with softness. Interestingly we show that for these systems with different degrees of fragility the relationship between the dynamics and softness is unique. Thus the richness of the theory is in its ability to explain this wide variety of systems. We also connect our work with machine learning studies. However, this study covers a wider range of systems and can be considered as a theoretical basis for the ML results explaining how and why the softness determines the dynamics. 

Our preliminary study shows that the present definition of softness when extended at a local level will allow us to identify the soft spots as has been done using other parametric descriptions of softness, including the ML studies \cite{richard,cubuk,Schoenholz_nature,Schoenholz263,sood_rajesh_nature}. Work in this direction is in progress.


\section{acknowledgement}
We thank Chandan Dasgupta and Olivier Dauchot for critically reading the manuscript. We thank Mohit Sharma and Vijayakumar Chikkadi for discussions. SMB thanks SERB for funding.


\end{document}


\title{A Microscopic Theory of Softness in Supercooled Liquids-Supplementary Information}
\author{Manoj Kumar Nandi}
\address{\textit{Polymer Science and Engineering Division, CSIR-National Chemical Laboratory, Pune-411008, India}}
\author{Sarika Maitra Bhattacharyya}
\email{mb.sarika@ncl.res.in}
\address{\textit{Polymer Science and Engineering Division, CSIR-National Chemical Laboratory, Pune-411008, India}}

\maketitle

\section{System details}
In this study, we perform molecular dynamics (MD) simulations for three-dimensional binary mixtures in the canonical ensemble. The system contains N number of particles within which the number of A type particles is $N_A$ and B type is $N_B$. In our simulations we have used periodic boundary conditions and Nos\'{e}-Hoover thermostat  with integration timestep 0.005$\tau$. The time
constants for  Nos\'{e}-Hoover thermostat  are taken to be 100  timesteps. The total number density $\rho=N/V$ is fixed, where V is the system volume. Length, temperature, and time are given in units of $\sigma_{AA}$ , $\epsilon_{AA}/k_B$,
$(m\sigma_{AA}^2 /\epsilon_{AA})^{1/2}$, respectively.
The models studied here are the binary Kob-Andersen Lennard-Jones (LJ) liquids \cite{kobandersenLJ} and its repulsive counterpart Weeks-Chandler-Andersen (WCA) model \cite{wca_model}, the Wahnstr{\"o}m model (WAHN) \cite{wahn_model} and the active system with different activities. We have carried out the molecular dynamics simulations (except for active systems) using the LAMMPS package \cite{lammps}. The data for active systems are obtained through private communication with R. Mandal, P. J. Bhuyan, M. Rao and C. Dasgupta. For all state points, three to five independent samples with run
lengths $>$ 100$\tau$ ($\tau_{\alpha}$ is the $\alpha$- relaxation time) are analyzed.

\subsection{LJ and WCA system}
The Kob-Andersen model (KA) which is the frequently studied glass forming model, is a binary mixture
(80:20) of Lennard-Jones (LJ) particles.
 The interatomic potential between the two species $\alpha$ and $\beta$, where $\alpha,\beta=A,B$, is described using truncated and shifted Lennard-Jones potential by;
\begin{equation}
 U_{\alpha\beta}(r)=
\begin{cases}
 U_{\alpha\beta}^{(LJ)}(r;\sigma_{\alpha\beta},\epsilon_{\alpha\beta})- U_{\alpha\beta}^{(LJ)}(r^{(c)}_{\alpha\beta};\sigma_{\alpha\beta},\epsilon_{\alpha\beta}),    & r\leq r^{(c)}_{\alpha\beta}\\
   0,                                                                                       & r> r^{(c)}_{\alpha\beta}
\end{cases}
\label{kalj_eq}
\end{equation}
where $U_{\alpha\beta}^{(LJ)}(r;\sigma_{\alpha\beta},\epsilon_{\alpha\beta})=4\epsilon_{\alpha\beta}[({\sigma_{\alpha\beta}}/{r})^{12}-({\sigma_{\alpha\beta}}/{r})^{6}]$ and
 $r^{(c)}_{\alpha\beta}=2.5\sigma_{\alpha\beta}$ for the LJ systems and $r^{(c)}_{\alpha\beta}=2^{1/6}\sigma_{\alpha\beta}$  for the WCA systems.
The interaction parameters for the systems are $\sigma_{AA}$ = 1.0, $\sigma_{AB}$ =0.8 ,$\sigma_{BB}$ =0.88,  $\epsilon_{AA}$ =1, $\epsilon_{AB}$ =1.5,
 $\epsilon_{BB}$ =0.5, $m_{A}$ = $m_B$=1.0. The densities we studied here are $\rho=1.2$ and 1.6 for the LJ model and $\rho =1.05$, 1.2 and 1.6 for the WCA model. The total number of particles N=1000 ($N_A=800$ and $N_B= 200$).
 \subsection{WAHN}
 The WHAN model is a binary (50:50) mixture of particles with the same interaction potential as described in Eq.\ref{kalj_eq} and the cutoff $r^{(c)}_{\alpha\beta}=2.5\sigma_{\alpha\beta}$. The interaction parameters are  $\sigma_{AA}$ = 1.0, $\sigma_{BB}$ =10/12, $\sigma_{AB}$ =($\sigma_{AA}+\sigma_{AB}$)/2. The interaction strength and masses are defined as  $\epsilon_{AA}$ = $\epsilon_{AB}$ = $\epsilon_{BB}$ =1.0 and $m_{A}$ =1, $m_B$=0.5. The  density for the system is $\rho=1.2959$ and the total number of particles N=500 ($N_A=250$ and $N_B=250$).

\subsection{Active systems}
As mentioned in the reference of the main paper the data of the active systems were not generated by us but obtained in private communication. The system details for active systems are given in Ref. [\onlinecite{active_system}]. However for sake of completeness of the present article, we give some brief details. The active systems, are the KA system interacting via Lennard-Jones potential at  $\rho=1.2$ with activity $f_0 = 0.5, 1.00, 1.75$,  where $f_{0}$ denotes the value of the self-propulsion force.
\section{Definitions}
\subsection{Relaxation times}
The $\alpha$-relaxation times are obtained from the decay of the self overlap function Q(t) using the
definition $Q(t=\tau_\alpha)=1/e$. The self overlap function Q(t) is a normalized two point correlation function and is defined as \cite{overlap_function},
\begin{eqnarray}
Q(t)= \Big< \frac{1}{N}\sum_{i=1}^N w(|\vec{r}_i(t)-\vec{r}_i(0)|)\Big>.
\end{eqnarray}
The window function $w(x)$ defines the condition of overlap between two particle positions separated by a time interval t and $w(x)$ is 1 when $x<a$ and zero otherwise.
The time dependent overlap function thus depends on
the choice of the cut-off parameter `a', which we choose
to be 0.3. This parameter is chosen such that particle
positions separated due to small amplitude vibrational
motion are treated as the same, or that $a^2$ is comparable to the value of the mean squared displacement (MSD) in the plateau between the ballistic and diffusive regimes.


\subsection{Static structure factor}
The partial structure factors $S_{\alpha\beta}({\bf q})$, needed as input for
the theory, can be defined as
\begin{equation}
    S_{\alpha\beta}({\bf q})=\frac{1}{\sqrt{N_{\alpha}N_\beta}}\sum_{i=1}^{N_\alpha}\sum_{j=1}^{N_\beta} exp[-i{\bf q}.({\bf r}_i^{\alpha}-{\bf r}_j^{\beta})].
    \label{struc_sq}
\end{equation}

\section{Details of the theory}
In the spirit of mean-field theory we can map the dynamics of a set of  interacting particles in terms of dynamics of a set of non-interacting particles in a potential $\Phi(r)$ as \cite{role_pair_correlation},
\begin{eqnarray}
 \frac{\partial \rho({\bf{r}},t)}{\partial t}={\bf{\nabla}}.\Big[\frac{k_BT}{\zeta}{\bf{\nabla}} \rho({\bf{r}},t)+\frac{\rho({\bf{r}},t)}{\zeta}{\bf{\nabla}} \Phi(r)\Big],
  \label{sml_eq}
\end{eqnarray}
Where $\rho(r,t)$ is the time dependent average density at "r" and $\Phi(r)$ is the effective potential felt by a particle at 'r' due to the interaction with the other particles present in the system. $\zeta$ is the value of friction. 
\begin{figure}[ht]
	\centering

		\includegraphics[width=0.44\textwidth]{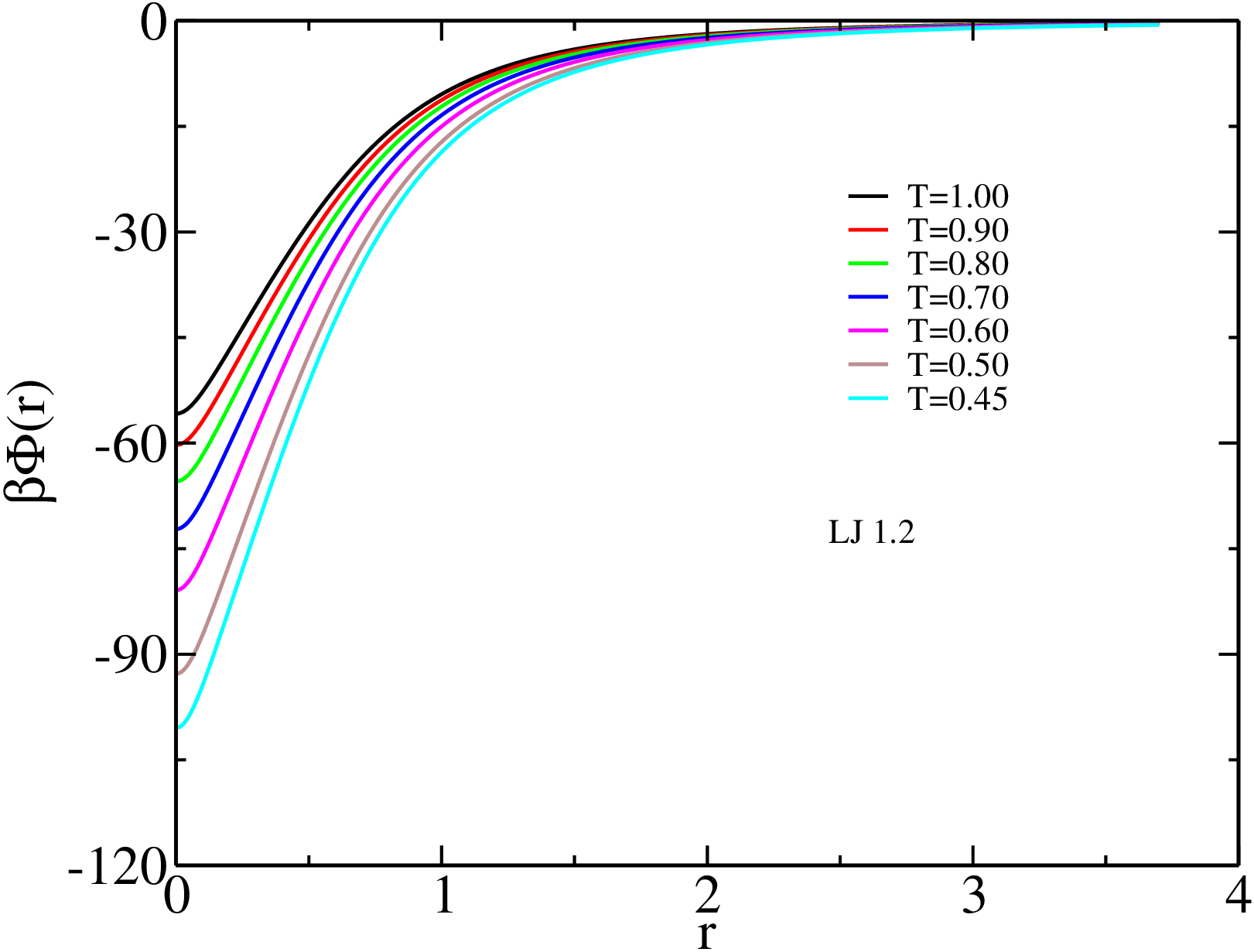}

	\caption{\it{  Temperature dependence of the potential obtained from Eq.\ref{mfpt_pot_binary}. As the temperature is lowered the potential becomes deeper.}}
	\label{fig.1}
\end{figure}
Using the dynamic density functional theory (DDFT) approach one writes the potential as $\Phi(r)={\delta F_{ex}[\rho(r)]/\delta \rho(r)}$ where $F_{ex}$ is the free energy. Next we use the Ramakrishnan-Yussouff (RY) excess free energy functional to derive the potential \cite{ry_form}. The form of $F_{ex}$ considering only up to pair interaction is given by,
\begin{eqnarray}
\beta F_{ex}(\rho({\bf R}))&\simeq& -\frac{1}{2}\int d{\bf R}\int d{\bf R'}\rho({\bf R},t)C(|{\bf R-R'}|)\rho({\bf R'},t)\nonumber\\
&=&-\frac{1}{2}\int \frac{d{\bf q}}{(2\pi)^3}C(q)\rho(-{\bf q},t)\rho({\bf q},t).
 \label{ry}
\end{eqnarray}
Where $\rho({\bf R},t)=<\sum_i\delta({\bf R-R}_i(t))>$ is the density and $C(|{\bf R-R'}|)$ is the direct correlation function, ${\bf R}$ being the position of the particle. Note that Eq.\ref{ry} depends on Eq.\ref{sml_eq} and both these two equations can be solved using an iterative method. \cite{archer2007,hopkins2010}. In an earlier work we have shown that Eq.\ref{sml_eq} can also be solved using some standard approximations \cite{krikpatrick1987,schweizer2005,role_pair_correlation,indranil}, which allow us to write the free energy in terms of the structure factor of the liquid and a displacement parameter. 

\begin{figure*}[ht]
	\centering
	\subfigure{
		\includegraphics[width=0.43\textwidth]{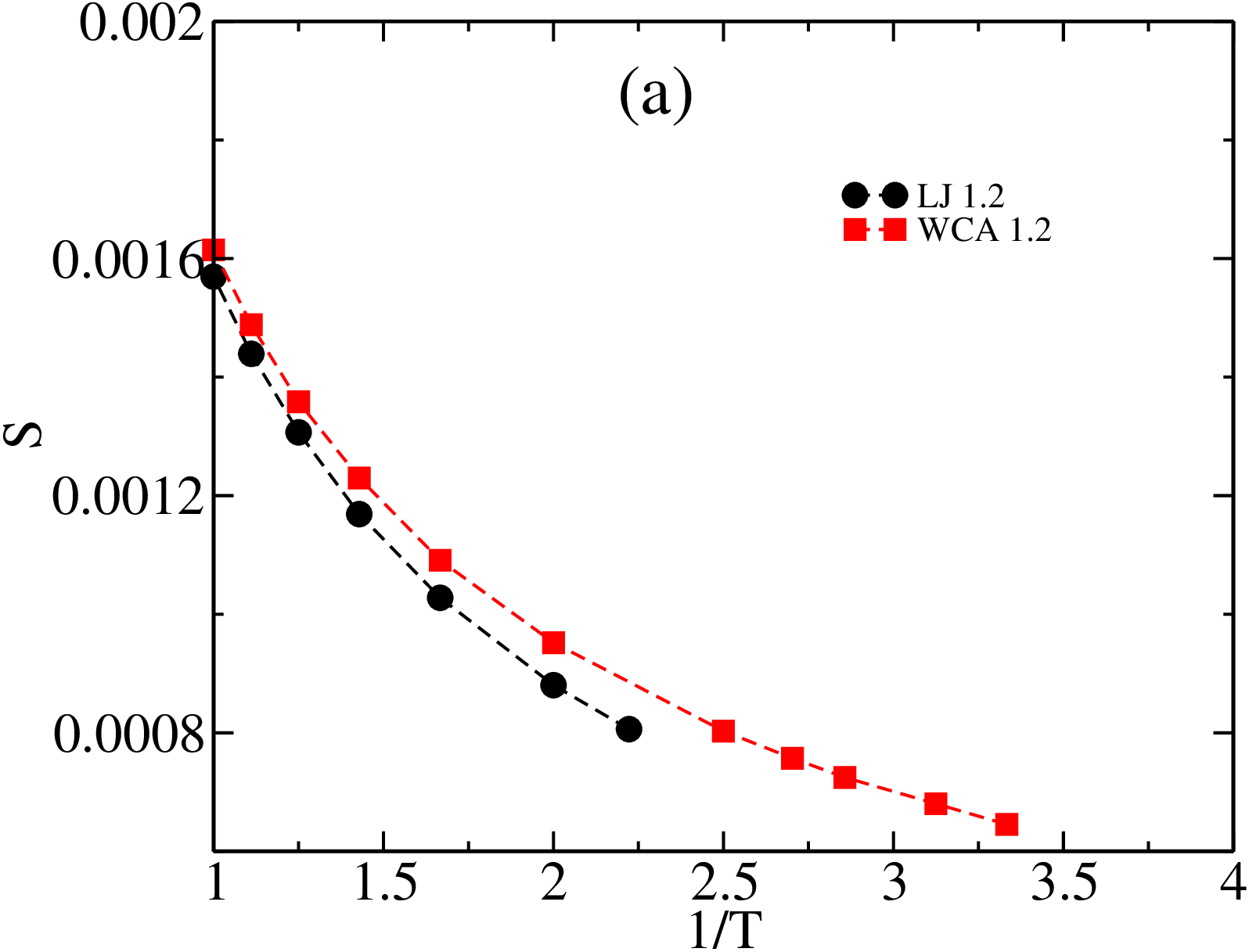}}
			\subfigure{
		\includegraphics[width=0.44\textwidth]{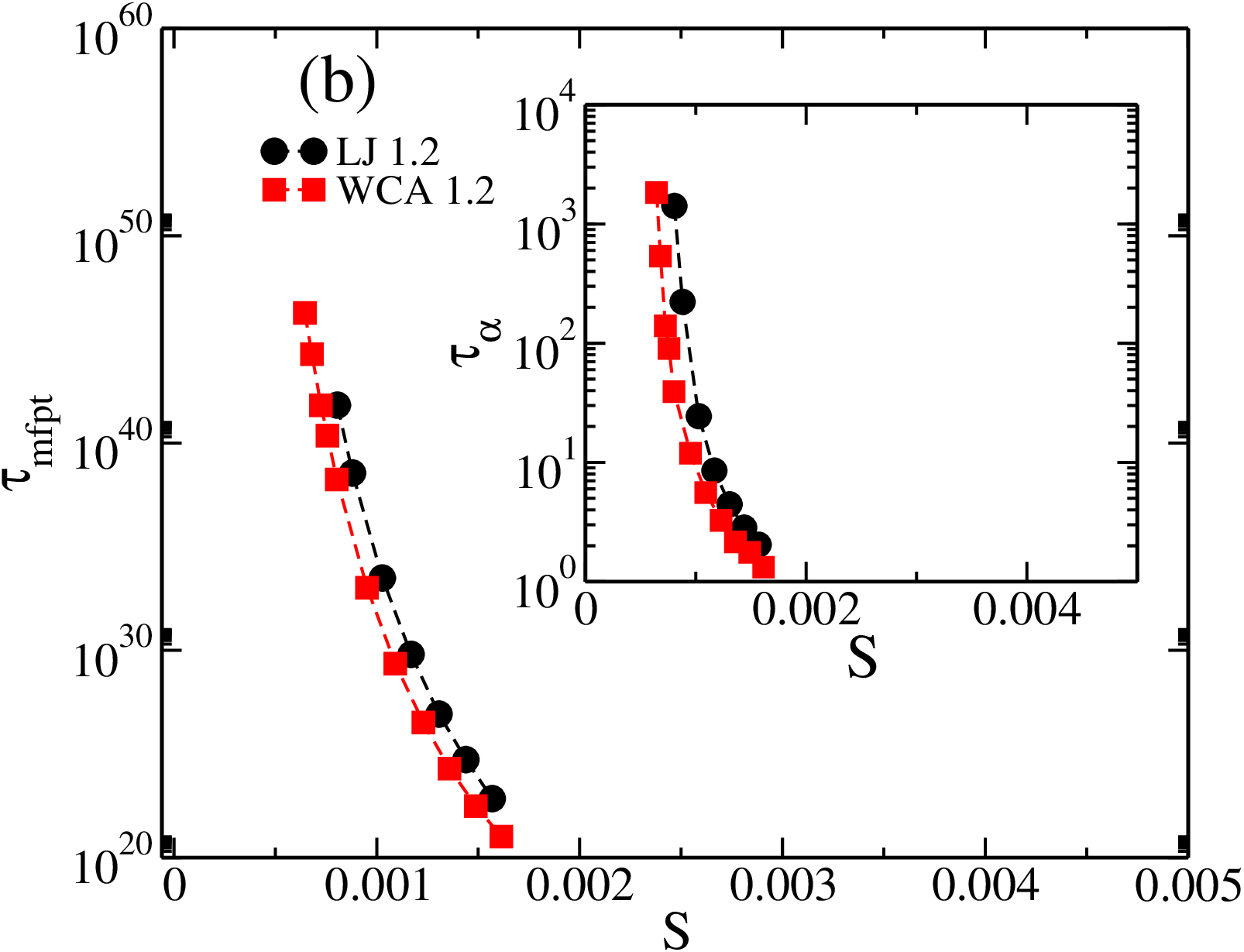}}
	\subfigure{
		\includegraphics[width=0.44\textwidth]{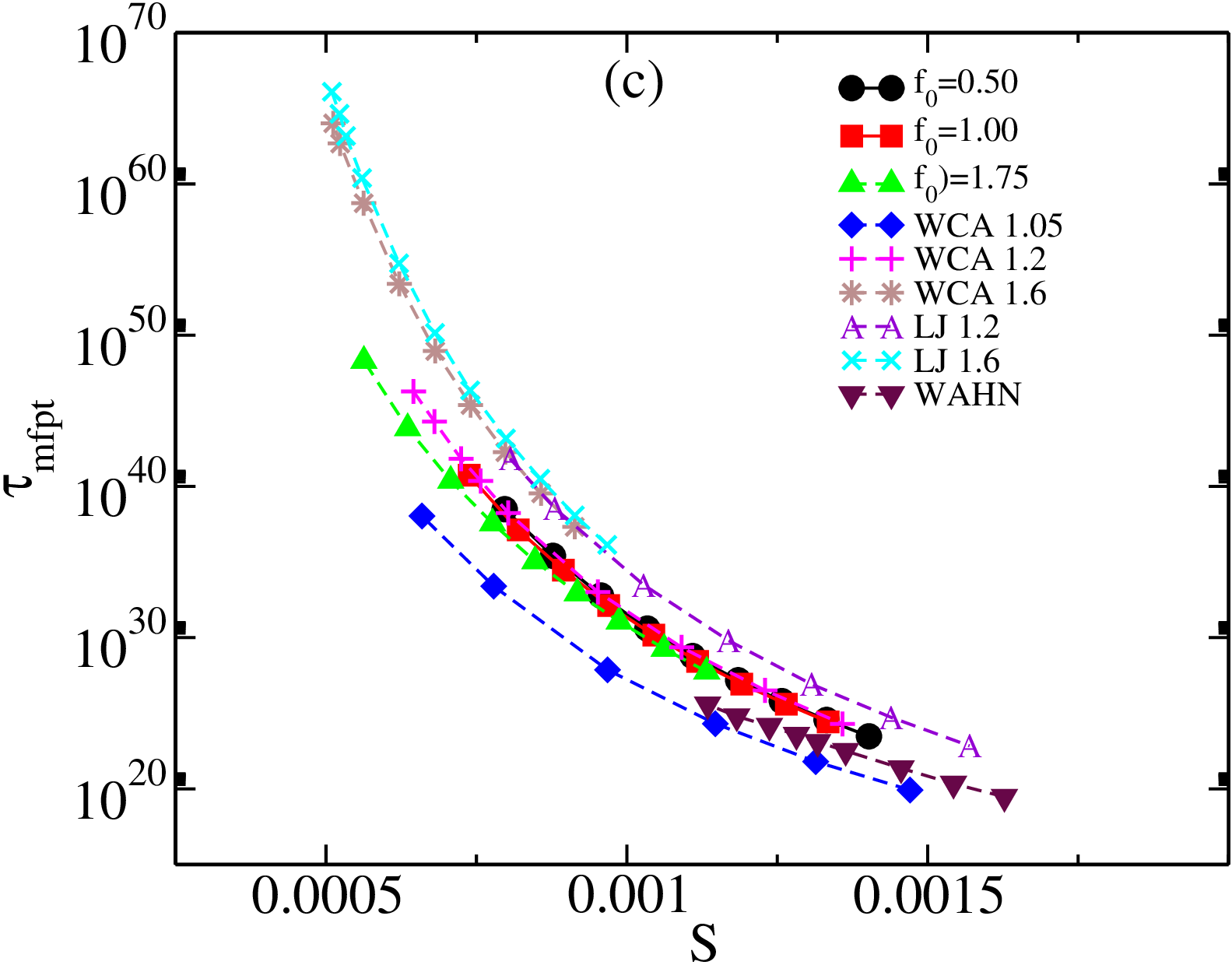}}

	\caption{\it{ (a) Softness (S) as a function of inverse temperature for LJ and WCA system at $\rho=1.2$. (b) $\tau_{mfpt}$, as a function of softness obtained from the harmonic fit of the potential for LJ and WCA systems with $\rho=1.2$. Inset: $\tau_{\alpha}$ against softness. Both the system shows similar behaviour with respect to softness parameter. (c) $\tau_{mfpt}$ against softness S for all the systems studied here. The plot does not show a master plot.}}
	\label{softness_fig2}
\end{figure*} 
First we make the Vineyard approximation with the assumption that both the single particle and the collective dynamics are similar which allow us to write $\rho({\bf{R}},t)=\rho({\bf{R}}(0))\rho_s({\bf{R}},t)$ and then to dynamically close the equation at single particle level we use $\rho^s(R,t)\simeq[3/2\pi <r^2_i(t)>]^{3/2}e^{[-3R^2/2<r_i^2(t)>]}$, and thus $\rho^s_q=e^{[-q^2<r_i^2(t)>/6]}$, where $<r_i^2(t)>$ is the mean square displacement of particle 'i'. We then drop the particle index 'i' and ignore the self-term ($i=j$) \cite{krikpatrick1987,schweizer2003,schweizer2005} in Eq.\ref{ry} as we are interested only in the effective interaction due to the presence of other particles. As mentioned before, within this framework each particle is independent and moving in a static field. Thus in the body fixed frame initially the particle is at the origin and the instantaneous position becomes its displacement. This allows us to write $<r^2(t)>=r^2(t)$ and the form of the approximate potential becomes, 
\begin{eqnarray}
 \beta \Phi(r)=-\int \frac{d{\bf{q}}}{(2\pi)^3} C(q)(S(q)-1)e^{-q^2r^2/6}.
\label{mfpt_pot}
\end{eqnarray}
Here S(q) is the structure factor of the liquid and $C(q)$ is the direct correlation function. For the binary system the potential becomes,
\begin{eqnarray}
\beta \Phi(r)=-\int \frac{d{\bf{q}}}{(2\pi)^3}\sum_{\alpha\beta} C_{\alpha\beta}(q)\sqrt{x_\alpha x_\beta}(S_{\alpha\beta}(q)-\delta_{\alpha\beta})e^{-q^2r^2/6}.
\label{mfpt_pot_binary}
\end{eqnarray}
Where $x_{\alpha/\beta}$ represents the fraction of particles of type $\alpha/\beta$ in the binary mixture. In the above expression $S_{\alpha\beta}(q)$ is obtained from Eq. \ref{struc_sq} and ${\bf{C(q)=1-S^{-1}(q)}}$, where ${\bf{C}}$, ${\bf{1}}$ and ${\bf{S(q)}}$ are in matrix form (For the derivation of binary form of the potential see  supplementary of Ref.\cite{role_pair_correlation}and Ref.\cite{indranil}).

\subsection {mfpt to Kramers dynamics}
\begin{eqnarray}
 \tau_{mfpt}=\frac{1}{D_0}\int_0^{r_{max}} e^{\beta\Phi(y)}dy\int_0^ye^{-\beta\Phi(z)}dz.
 \label{taumfpt}
\end{eqnarray}

 In our earlier study \cite{indranil}, we have shown that in certain limit Eq.\ref{taumfpt} reduces to the expression of Kramers barrier crossing dynamics \cite{zwanzig}. Here we sketch the steps to arrive at the Kramers expression. When $k_BT$ is small, the z integration in the $\tau_{mfpt}$ calculation is dominated by the minimum and the y integration is dominated by the barrier of the potential. For z integral, we can expand the potential $\Phi(z)$ up to quadratic order $\Phi(z)=\Phi_{min}+\frac{1}{2}\omega^2_{min}(z-z_0)^2+...$ and we can write the upper limit of the integration as infinity. Similarly for y integral we can expand $\Phi(y)=\Phi_{max}-\frac{1}{2}\omega^2_{max}(y-y_0)^2+...$ and following similar logic we can write $r_{max}$ as infinity. With these approximations the equation for $\tau_{mfpt}$ becomes, 

\begin{eqnarray}
\tau_{mfpt}&=&\frac{1}{D_0}\Big(\frac{1}{2}\sqrt{\frac{2\pi K_BT}{\omega_{max}^2}}e^{\beta \Phi_{max}}\Big)\Big(\frac{1}{2}\sqrt{\frac{2\pi K_BT}{\omega_{min}^2}}e^{-\beta \Phi_{min}}\Big)\nonumber\\
&=&\Big(\frac{\pi K_BT}{2D_0\omega_{max}\omega_{min}}\Big)e^{\beta(\Phi_{max}-\Phi_{min})}\nonumber\\
&=&\tau_0e^{\beta\Delta \Phi(T)}=\tau_{Kramers}
\label{kramers}
\end{eqnarray}
\noindent

\begin{table}[h]
\caption{\textit{{\bf{Fragility of the systems}}:The VFT parameters when fitted against temperature. [$\tau_{\alpha}=\tau_0 exp(1/K_{VFT}(T/T_{VFT}-1)$].}}
\begin{center}
 \begin{tabular}{||c |c | c||} 
  \hline
systems & $K_{VFT}$ & $T_{VFT}$  \\  
 \hline\hline
   active $f_0$ 0.50 & 0.200 & 0.275  \\
 \hline
 active $f_0$ 1.00 & 0.156 & 0.215 \\
 \hline
active $f_0$ 1.75 & $10^{-19}$ & $10^{-19}$  \\
 \hline
  WCA 1.05 & $10^{-19}$ & $10^{-19}$ \\
 \hline
 WCA 1.2 & 0.167 & 0.174 \\
 \hline
  WCA 1.6 & 0.239 & 1.185\\
 \hline
 LJ 1.2 & 0.189 & 0.279 \\
 \hline
 LJ 1.6 & 0.317 & 1.332 \\
 \hline
   WAHN & 0.656 & 0.476 \\
 \hline

\end{tabular}
\end{center}
\end{table}

\subsection{Temperature dependence of potential energy and softness}
In Fig.\ref{fig.1} we plot the potential at different temperatures and show that as the temperature is lowered, the potential becomes deeper leading to stronger caging of the particle. The plot shows that an increase in the depth of the potential also leads to a decrease in its softness. To quantify the softness of the potential we fit the potential near r=0 to a harmonic form, $\Phi(r)=\Phi(r=0)+\frac{1}{2}r^2/S$, where $\Phi(r=0)$ is the value of the potential at the minima and the softness of the potential as $S$. In Fig.\ref{softness_fig2} we plot the softness of the potential against the inverse of temperature for LJ and WCA systems at density 1.2. We find that at the same temperature the WCA system is softer than the LJ system and with a decrease in temperature this difference in softness increases (Fig.\ref{softness_fig2}(a)). 
To understand the effect of this softness on the dynamics, in Fig.\ref{softness_fig2}(b) we plot the $\tau_{mfpt}$ against softness for LJ and WCA systems. The two plots almost overlap. In the inset we plot the $\alpha $ relaxation time, $\tau_{\alpha}$ against softness which also shows a master plot. These observations are  similar to those reported in the machine learning study \cite{atractive_truncated_Landes}. However when we plot the mfpt relaxation time of all the systems studied here against their respective S values we see that there is a spread in the data.(Fig.\ref{softness_fig2}(c)).

\begin{figure}[ht]
\includegraphics[width=0.44\textwidth]{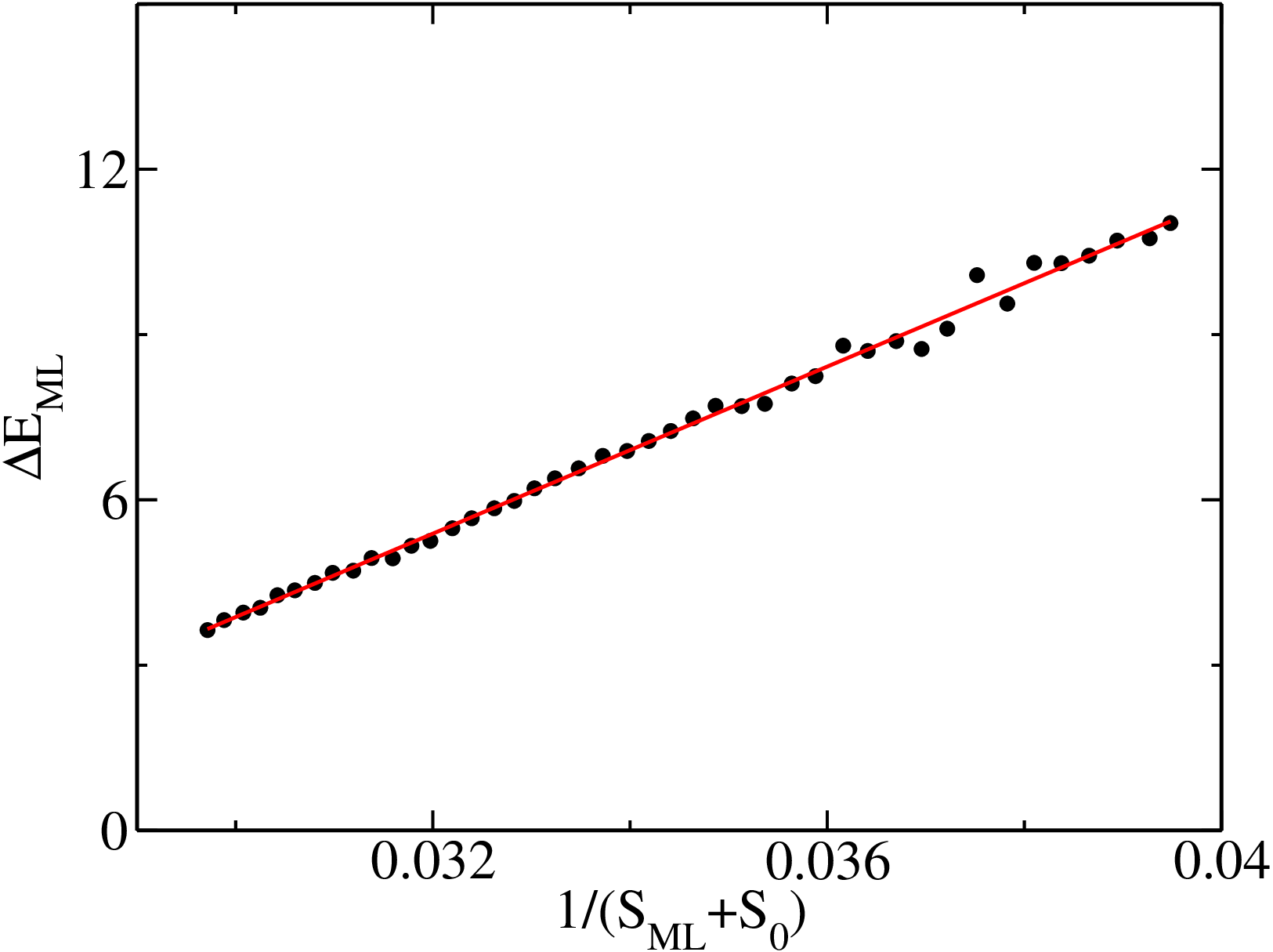}
\caption{\it{$\Delta E_{ML}$ and $S_{ML}$ from Ref\cite{Schoenholz_nature} (Fig.7), re-plotted against $1/(S_{ML}+S_{0})$ where $S_{0}=29.92\pm 1.70$ is a fitting constant.}}
\label{ML_MF}
\end{figure}

We  show  in  the  main  paper  that  the  softness  obtained  via  ML  studies  is  linearly  proportional  to  that obtained in the present study.  However, in our study we  find  that  the  energy  barrier  related  to  the  dynamics  is  inversely  proportional  to  S  whereas  in  the  ML study it is linearly proportional to S.  In  the  supplementary  of  Ref \cite{Schoenholz_nature} (Fig.7)  the  authors show that the softness dependence of $\Delta E_{ML}$ is not really linear but quadratic.  Here we show that $\Delta E_{ML}$ vs $1/(S_{ML}+S_0)$  shows  a  linear  behaviour  where $S_{0}$ is  a fitting constant ( Fig.3).  A Taylor’s series expansion of $1/(S_{ML}+S_0) =\frac{1}{S_0}-\frac{S_{ML}}{S^2_0}+\frac{S^2_{ML}}{S^3_0}-.... $ This expansion although gives quadratic terms the high value of $S_0$ reduces the weightage of the quadratic term.   Hence the earlier described quadratic dependence of $\Delta E_{ML}$  with $S_{ML}$ is similar to the dependence of $\Delta E_{ML}$ against an inverse softness shifted by $S_{0}$.

\begin{table}[h]
\caption{\textit{The fitting parameters when $Tln\tau_{\alpha}$ fitted against 1/S. [$Tln\tau_{\alpha}=C'+B'/S$].}}
\begin{center}
 \begin{tabular}{||c |c | c||} 
  \hline
systems & $C'$ & $B'$  \\  
 \hline\hline
   active $f_0$ 0.50 & -4.33096 & 0.00494442  \\
 \hline
 active $f_0$ 1.00 & -3.98141 & 0.00391527 \\
 \hline
active $f_0$ 1.75 & -0.583467 & 0.000759566  \\
 \hline
  WCA 1.05 & 0.0128896 & 0.000253414 \\
 \hline
 WCA 1.2 & -1.95458 & 0.0020247 \\
 \hline
  WCA 1.6 & -13.5282 & 0.00760871\\
 \hline
 LJ 1.2 & -3.60636 & 0.0042633 \\
 \hline
 LJ 1.6 & -14.356 & 0.00852563 \\
 \hline
   WAHN & -5.92496 & 0.00730347 \\
 \hline
\end{tabular}
\end{center}
\end{table}
\begin{table}[h]
\caption{\textit{The fitting parameters when S fitted against T. [$S=a_0+a_1T$].}}
\begin{center}
 \begin{tabular}{||c |c | c||} 
  \hline
systems & $a_0$ & $a_1$  \\  
 \hline\hline
   active $f_0$ 0.50 & 0.000198276 & 0.00151242  \\
 \hline
 active $f_0$ 1.00 & 0.000218658 & 0.00149788 \\
 \hline
active $f_0$ 1.75 & 0.000280846 & 0.00141684  \\
 \hline
  WCA 1.05 & 0.000483458 & 0.00328831 \\
 \hline
 WCA 1.2 & 0.000215331 & 0.00145775 \\
 \hline
  WCA 1.6 & $8.82208\times 10^{-5}$ & 0.000236547\\
 \hline
 LJ 1.2 & 0.000176762 & 0.00140982 \\
 \hline
 LJ 1.6 & $8.76168\times 10^{-5}$ & 0.000236096 \\
 \hline
   WAHN & $8.03145\times10^{-5}$ & 0.00182588 \\
 \hline
\end{tabular}
\end{center}
\end{table}

